\documentclass[a4paper,twocolumn,showpacs]{revtex4}

\usepackage{amsmath,amssymb,amsfonts}
\usepackage{graphicx}
\usepackage{psfrag}
\usepackage{color}
\usepackage{hyperref}
\hypersetup{
	pdfauthor={C. Buzano, E. De Stefanis, and M. Pretti},
	pdftitle={Cluster-variation approximation for a network-forming lattice-fluid model},
	colorlinks,linkcolor={blue},citecolor={red}}

\newcommand{\ham}{\mathcal{H}}

\begin{document}

\title{Cluster-variation approximation for a network-forming lattice-fluid model}

\author{C.~Buzano}

\affiliation{Dipartimento di Fisica, Politecnico di Torino, Corso
Duca degli Abruzzi 24, I-10129 Torino, Italy}

\author{E.~{De~Stefanis}}

\affiliation{Dipartimento di Fisica, Politecnico di Torino, Corso
Duca degli Abruzzi 24, I-10129 Torino, Italy}

\affiliation{Center for Statistical Mechanics and Complexity,
CNR-INFM Roma 1, Piazzale Aldo Moro 2, I-00185 Roma, Italy}

\author{M.~Pretti}

\affiliation{Dipartimento di Fisica, Politecnico di Torino, Corso
Duca degli Abruzzi 24, I-10129 Torino, Italy}

\affiliation{Center for Statistical Mechanics and Complexity,
CNR-INFM Roma 1, Piazzale Aldo Moro 2, I-00185 Roma, Italy}

\date{\today}

\begin{abstract}
We consider a 3-dimensional lattice model of a network-forming
fluid, which has been recently investigated by Girardi and
coworkers by means of Monte Carlo simulations [J. Chem. Phys.
\textbf{126}, 064503 (2007)], with the aim of describing water
anomalies. We develop an approximate semi-analytical calculation,
based on a cluster-variation technique, which turns out to
reproduce almost quantitatively different thermodynamic properties
and phase transitions determined by the Monte Carlo method.
Nevertheless, our calculation points out the existence of two
different phases characterized by long-range orientational order,
and of critical transitions between them and to a high-temperature
orientationally-disordered phase. Also, the existence of such
critical lines allows us to explain certain ``kinks'' in the
isotherms and isobars determined by the Monte Carlo analysis. The
picture of the phase diagram becomes much more complex and richer,
though unfortunately less suitable to describe real water.
\end{abstract}

\pacs{
05.50.+q   
61.20.Gy   
65.20.-w   
}

\maketitle

\section{Introduction}

Water attracts great interest from physicists, due to its
different anomalous
properties~\cite{EisenbergKauzmann1969,Franks1982,Stanley2003}.
Just to mention a few of them, it is well known that, at ordinary
pressures, the solid phase (ice) is less dense than the
corresponding liquid phase, and the latter displays a temperature
of maximum density, just above the freezing transition.
Furthermore, the heat capacity of liquid water is unusually large,
whereas both isothermal compressibility and isobaric heat capacity
display a minimum as a function of temperature. Although a full
prediction of such and other anomalies from first principles has
not been given yet, it is widely believed that it should be
related to the ability of water molecules to form a network of
hydrogen bonds.

Among different possible approaches, a substantial body of
theoretical investigations concerning so-called network-forming
fluids has been developed in the framework of lattice
models~\cite{BellLavisI1970,BellLavisII1970,SouthernLavis1980,HuckabyHanna1987,PatrykiejewPizioSokolowski1999,BuzanoDestefanisPelizzolaPretti2004,Bell1972,BellSalt1976,WilsonBell1978,MeijerKikuchiPapon1981,MeijerKikuchiVanRoyen1982,LavisSouthern1984,BesselingLyklema1994,BesselingLyklema1997,RobertsDebenedetti1996,RobertsPanagiotopoulosDebenedetti1996,RobertsKarayiannakisDebenedetti1998,PrettiBuzano2004,PrettiBuzano2005,GirardiBalladaresHenriquesBarbosa2007,GirardiSzortykaBarbosa2007,SastrySciortinoStanley1993jcp,BorickDebenedettiSastry1995,SastryDebenedettiSciortinoStanley1996,RebeloDebenedettiSastry1998}.
One can use various types of model molecules, characterized by
orientation-dependent interactions, in either
two~\cite{BellLavisI1970,BellLavisII1970,SouthernLavis1980,HuckabyHanna1987,PatrykiejewPizioSokolowski1999,BuzanoDestefanisPelizzolaPretti2004}
or
three~\cite{Bell1972,BellSalt1976,WilsonBell1978,MeijerKikuchiPapon1981,MeijerKikuchiVanRoyen1982,LavisSouthern1984,BesselingLyklema1994,BesselingLyklema1997,RobertsDebenedetti1996,RobertsPanagiotopoulosDebenedetti1996,RobertsKarayiannakisDebenedetti1998,PrettiBuzano2004,PrettiBuzano2005,GirardiBalladaresHenriquesBarbosa2007,GirardiSzortykaBarbosa2007,SastrySciortinoStanley1993jcp,BorickDebenedettiSastry1995,SastryDebenedettiSciortinoStanley1996,RebeloDebenedettiSastry1998}
dimensions. A natural choice for water appears to be a
3-dimensional model molecule with four bonding arms arranged in a
tetrahedral
symmetry~\cite{Bell1972,BellSalt1976,WilsonBell1978,MeijerKikuchiPapon1981,MeijerKikuchiVanRoyen1982,LavisSouthern1984,BesselingLyklema1994,BesselingLyklema1997,RobertsDebenedetti1996,RobertsPanagiotopoulosDebenedetti1996,RobertsKarayiannakisDebenedetti1998,PrettiBuzano2004,PrettiBuzano2005,GirardiBalladaresHenriquesBarbosa2007,GirardiSzortykaBarbosa2007}.
Two of them can represent the hydrogen (H) atoms, which are
positively charged and act as donors for the H bond, whereas the
other two can represent the negatively charged regions (``lone
pairs'') present in the $\mathrm{H}_2\mathrm{O}$ molecule, acting
as H bond acceptors. As far as the lattice is concerned, the
body-centered cubic (bcc) lattice is suitable for the tetrahedral
molecule, as the latter can point its arms towards four out of
eight nearest neighbors of each given site. The above features are
common to several models, which differ in the form of interactions
and allowed configurations.

In the early model proposed by
Bell~\cite{Bell1972,BellSalt1976,WilsonBell1978}, molecules can
point their arms only to nearest neighbor sites. An attractive
energy is assigned for every pair of occupied nearest neighbors,
with an extra contribution if a hydrogen bond is formed, i.e., if
a donor arm points to an acceptor arm. Furthermore, a repulsive
energy is assigned for certain triplets of occupied sites, in
order to account for the difficulty of forming H bonds by closely
packed water molecules. Minor variations of the Bell model have
been proposed by Meijer and
coworkers~\cite{MeijerKikuchiPapon1981,MeijerKikuchiVanRoyen1982},
which replaced the three-body interaction with a simple
next-nearest-neighbor repulsion, and by Lavis and
Southern~\cite{LavisSouthern1984}, who defined a simplified model
with no distinction between donors and acceptors, first pointing
out that such a distinction is scarcely important for a
qualitative description of water thermodynamics. All these studies
are mainly focused on the equilibrium phase diagram of water, and
predict a liquid-vapor coexistence and two different
low-temperature phases characterized by long-range orientational
order and different densities. At zero temperature, the
low-density phase is an ideal diamond network made up of H bonded
water molecules, with half the lattice sites left empty,
resembling the structure of ice Ic (cubic ice). At the melting
temperature, this phase correctly displays a lower density than
the liquid phase. Conversely, the high-density phase at zero
temperature is made up of two interpenetrating diamond structures,
with all sites occupied, resembling the structure of ice VII (a
high pressure form of ice).

More recently, a similar model has been proposed by Roberts and
Debenedetti~\cite{RobertsDebenedetti1996,RobertsPanagiotopoulosDebenedetti1996,RobertsKarayiannakisDebenedetti1998},
with the aim of exploring the phase diagram of water deep into the
supercooled liquid region, and, more generally, the possibility of
liquid-liquid immiscibility in a network-forming fluid. There are
two differences with respect to the original Bell model. First,
water molecules have an additional number of configurations, in
which they cannot form H-bonds. Such a number is an adjustable
parameter, related to the entropy of breaking H bonds. Second, the
energy penalty for the triplets occurs only when two molecules in
a triplet form a H bond. At zero temperature, this model still
predicts the two ordered H-bond networks of the Bell model, but
these phases have never been investigated at finite temperature,
due to the main attention devoted by the cited papers to (stable
or metastable) liquid phases. In particular, these studies have
given a contribution in the framework of a debate between two
different conjectures put forth respectively by Speedy and
Angell~\cite{SpeedyAngell1976} and Poole and
coworkers~\cite{PooleSciortinoEssmannStanley1992} to explain
apparent power-law divergences in the response functions of
supercooled water~\cite{SpeedyAngell1976}. In a few words, Speedy
and Angell postulated a reentrance of the stability limit
(spinodal) of liquid water, whereas Poole and coworkers postulated
the existence of two different metastable liquid phases (in
analogy with the two different amorphous ices observed in
experiments), whose coexistence line could possibly end in a
(metastable) critical point. The model by Roberts and Debenedetti
turned out to be compatible with the second critical point
conjecture, which has also collected several other indirect
evidences, both in experiments~\cite{MishimaStanley1998} and
computer simulations~\cite{Harrington1997}. Two authors of the
present paper have studied a simplified version of the
Roberts-Debenedetti model, without the donor-acceptor
asymmetry~\cite{PrettiBuzano2004,PrettiBuzano2005}, pointing out
that the model is indeed compatible with both the aforementioned
conjectures (depending on parameter values), and is also able to
describe the anomalous properties of water as a solvent for apolar
molecules (hydrophobic effect).

A different variation of the Bell model has been considered by
Besseling and
Lyklema~\cite{BesselingLyklema1994,BesselingLyklema1997}, who have
taken into account only nearest-neighbor interactions, namely, an
attractive term for H-bonded molecules and a repulsive term for
nonbonded molecules. Even in this case, the two different ordered
phases described above are stable at zero temperature, but they
have not been investigated at finite temperature. The cited
studies were indeed focused on the liquid-vapor interface
properties~\cite{BesselingLyklema1994} and on hydrophobic
hydration thermodynamics~\cite{BesselingLyklema1997}, for which
the authors obtained good agreement with experiments. These
results were obtained in the so-called quasi-chemical or Bethe
approximation, i.e., a first-order approximation which takes into
account correlations over clusters made up of two nearest-neighbor
sites.

Very recently, Girardi and
coworkers~\cite{GirardiBalladaresHenriquesBarbosa2007} have
performed extensive Monte Carlo simulations for the previously
described model, in the simplified version obtained by removing
the donor-acceptor distinction. This work suggests the existence
of two liquid phases of different densities, which can coexist in
equilibrium, in agreement with the conjecture of Poole and
coworkers~\cite{PooleSciortinoEssmannStanley1992}. The coexistence
line terminates in a critical point, whereas the low-density phase
exhibits a temperature of maximum density (TMD), depending on
pressure. A subsequent work~\cite{GirardiSzortykaBarbosa2007}
suggests a relationship between the density anomaly and the
anomalous behavior of the diffusion coefficient.

In this paper we analyze the above model by a generalized
first-order approximation, based on a 4-site tetrahedral cluster.
Such a work is motivated by the fact that the reliability of
cluster-variational approximations, though widely employed for
this kind of models with orientation-dependent interactions, has
been scarcely tested against Monte Carlo
simulations~\cite{WhitehouseChristouNicholsonParsonage1984}. Let
us anticipate that our calculation reproduces most physical
properties obtained by simulations (namely, isotherms, isobars,
binodals, TMD locus) with remarkable accuracy. Nevertheless, the
picture of the phase diagram turns out to be quite different, as
the two different condensed phases exhibit long-range
orientational order, which lead us to identify them as a
continuous temperature evolution of the two zero-temperature
network structures introduced above. As a consequence, the two
critical points are indeed tricritical, and two critical lines
appear, indicating two different kinds of symmetry breaking,
respectively from an orientationally-disordered phase to the
high-density ordered phase and from the latter to the low-density
one. In order to check the existence of critical lines and
symmetry breaking we have performed some new Monte Carlo
simulations in specific regions of the phase diagram.

The paper is organized as follows. In Section~II we describe the
model hamiltonian and analyze the ground-state. In Sec.~III we
introduce the cluster-variational technique employed for the
calculation. Sec.~IV describes our results, discussing them in
comparison with those obtained from numerical simulations in
Ref.~\cite{GirardiBalladaresHenriquesBarbosa2007}. Sec.~V reports
on the new Monte Carlo results, while Sec.~VI is devoted to some
concluding remarks.

\section{Model and ground state}

\begin{figure}
  \includegraphics*[20mm,193mm][100mm,268mm]{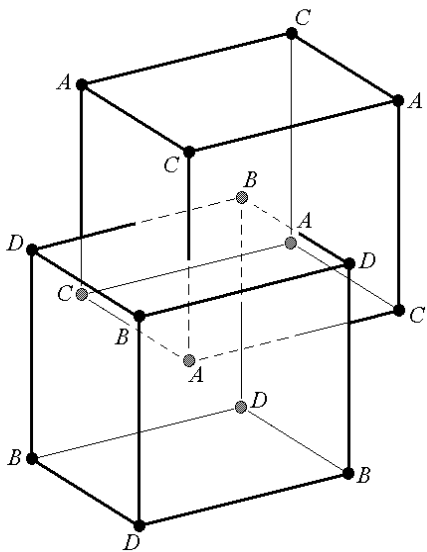}
  \caption
  {
    Two conventional (cubic) cells of the body centered cubic (bcc) lattice.
    $A,B,C,D$ denote four interpenetrating face-centered cubic (fcc) sublattices.
  }
  \label{fig:reticolo}
\end{figure}

\begin{figure}
  \includegraphics*[20mm,193mm][100mm,268mm]{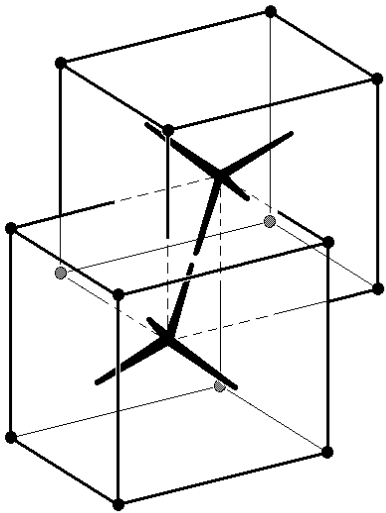}
  \caption
  {
    Two model molecules forming a H~bond.
    The lower molecule is in the $i=1$ configuration,
    the upper one is in the $i=2$ configuration (see the text).
  }
  \label{fig:molecole}
\end{figure}

As mentioned in the Introduction, the model is defined on a
body-centered cubic lattice (see Fig.~\ref{fig:reticolo}).
Each site can be empty or occupied by a molecule. The model
molecule possesses the tetrahedral structure introduced above,
with 4 equivalent bonding arms, which can point towards 4 out of 8
nearest neighbors of a given site. A hydrogen bond is formed,
yielding an attractive energy $-\eta < 0$, whenever two
nearest-neighbor molecules point an arm to each other, with no
distinction between donor and acceptor (see
Fig.~\ref{fig:molecole}).
Moreover, a repulsive interaction $\epsilon > 0$ is assigned to
any pair of nearest-neighbor sites occupied by water molecules.
This choice of the interaction parameters implies an energetic
penalty for nearest-neighbor molecules not forming H bonds.
Finally, as we find it convenient to work in the grand-canonical
ensemble, a chemical potential contribution $-\mu$ is taken into
account for every occupied site.

\begin{figure}
  \resizebox{80mm}{!}{\includegraphics*[20mm,195mm][90mm,277mm]{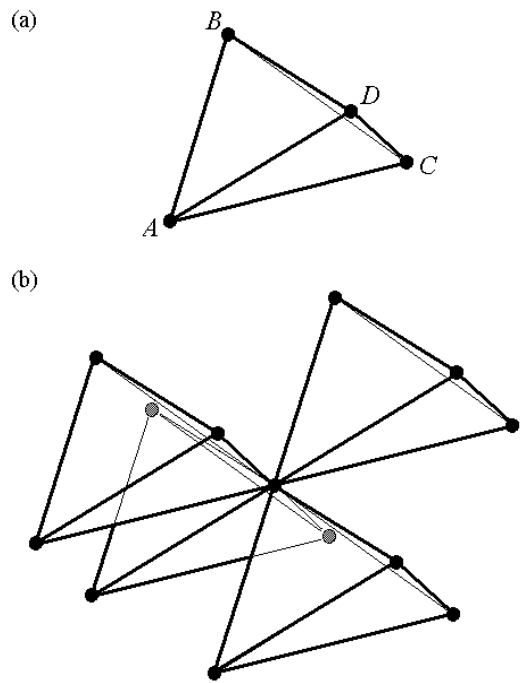}}
  \caption
  {
    (a) Basic cluster (irregular tetrahedron):
    $A,B,C,D$ denote sites in the 4 corresponding sublattices. $AB$,
    $BC$, $CD$, and $DA$ are nearest-neighbor pairs;
    $AC$ and $BD$ are next-nearest-neighbor pairs.
    (b) Husimi tree structure corresponding to the generalized first-order
    approximation on the tetrahedron.
  }
  \label{fig:cactustetraedro}
\end{figure}

In principle, the hamiltonian of the system can be written as a
sum of coupling terms between nearest-neighbor sites.
Nevertheless, with a view to subsequent analytical development,
let us write the hamiltonian as a sum over irregular tetrahedral
clusters, whose vertices lie on 4 different face-centered cubic
sublattices [see Figs.~\ref{fig:reticolo}
and~\ref{fig:cactustetraedro}(a)]. Let us note that there are 24
such tetrahedra sharing a given site, but it is sufficient to
choose only a subset of 4, in order to cover all the
nearest-neighbor pairs [Fig.~\ref{fig:cactustetraedro}(b)]. There
are indeed 6 different possibilities to choose a proper subset,
but the hamiltonian turns out to be independent of this choice.
We thus finally write
\begin{equation}
  \ham =
  \sum_{\left( \alpha,\beta,\gamma,\delta \right)}
  \ham_{i_\alpha i_\beta i_\gamma i_\delta}
  ,
  \label{eq:ham}
\end{equation}
where the elementary contribution $\ham_{ijkl}$ will be denoted as
tetrahedron hamiltonian. The subscripts
$i_\alpha,i_\beta,i_\gamma,i_\delta$ denote site configurations
for the 4 vertices $\alpha,\beta,\gamma,\delta$ of each
tetrahedron. Possible site configurations are as follows: $i=0$
denotes a vacancy (empty site), whereas $i=1,2$ denote a molecule
in its 2 possible orientations (see Fig.~\ref{fig:molecole}). Due
to the presence of only nearest-neighbor interactions, and
assuming that $(i,j)$, $(j,k)$, $(k,l)$, and $(l,i)$ just refer to
nearest-neighbor configurations, the tetrahedron hamiltonian can
be written as
\begin{equation}
  \ham_{ijkl} = \ham_{ij} + \ham_{jk} + \ham_{kl} + \ham_{li}
  ,
  \label{eq:tetraham}
\end{equation}
where
\begin{equation}
  \ham_{ij} = \epsilon n_i n_j - \eta h_{ij} - \mu n_i/4
  .
\end{equation}
In the latter equation, $n_i$ is an occupation variable, defined
as $n_i=0$ if $i=0$ (empty site), $n_i=1$ otherwise (occupied
site), whereas $h_{ij}$ is a bond variable, defined as $h_{ij}=1$
if the pair configuration $(i,j)$ represents a H~bond, and
$h_{ij}=0$ otherwise. Let us also assume that $i,j,k,l$ (in this
order) denote configurations of sites placed on, say, $A,B,C,D$
sublattices respectively. If $A,B,C,D$ are defined as in
Fig.~\ref{fig:reticolo}, we can define $h_{ij}=1$ if $i=1$ and
$j=2$, and $h_{ij}=0$ otherwise. The $1/4$ coefficient is meant to
avoid multiple countings of the chemical potential terms.

Let us now define the tetrahedron configuration probability by
$p_{ijkl}$, with the same convention about the subscript order,
and assume that the probability distribution is equal for every
tetrahedron. Such an assumption seems to be reasonable, as it
allows one to characterize the two ordered structures present in
the ground state. In both cases, the distribution is a delta
function concentrated on a fixed tetrahedron configuration
$(i,j,k,l)$ (i.e., $p_{ijkl} = 1$ for that configuration, and $0$
otherwise), so that there is no residual entropy. As far as the
low-density (LD) structure is concerned, it indeed turns out to be
four-fold degenerate, since it can be represented by four
alternative tetrahedron configurations, namely,
$(i,j,k,l)=(1,2,0,0),(0,1,2,0),(0,0,1,2),(2,0,0,1)$ (obtained from
one another by a circular permutation). In each structure, a pair
of sublattices (respectively, $AB$, $BC$, $CD$, and $DA$) is
occupied by fully bonded molecules (forming a diamond structure),
while the other two sublattices are empty. Conversely, the
high-density (HD) structure turns out to be two-fold degenerate,
being represented by the two alternative configurations
$(i,j,k,l)=(1,2,1,2),(2,1,2,1)$. Both these configurations have
all lattice sites occupied, but they differ in the pairs of fully
bonded sublattices, that is, $AB$ and $CD$ in the former case and
$BC$ and $DA$ in the latter.

The grand-canonical energy per site in the thermodynamic limit can
be written as
\begin{equation}
  w = \sum_{i,j,k,l=0}^2  p_{ijkl} \ham_{ijkl} ,
  \label{eq:intenergy}
\end{equation}
which takes into account that there is one tetrahedron per site.
By grand-canonical energy, we mean
\begin{equation}
  w = u - \mu \rho ,
\end{equation}
where $u$ is the internal energy per site and $\rho$ is the
density, i.e., the average occupation probability
\begin{equation}
  \rho = \sum_{i,j,k,l=0}^2  p_{ijkl} \frac{n_i+n_j+n_k+n_l}{4} .
  \label{eq:density}
\end{equation}

We are now in a position to evaluate the energy of the two
ground-state structures described above. For the LD phase, we have
to replace
\begin{equation}
  p_{ijkl} = \delta_{i,1} \delta_{j,2} \delta_{k,0} \delta_{l,0}
\end{equation}
into Eq.~\eqref{eq:intenergy}, obtaining
\begin{equation}
  w_{\text{LD}} = \epsilon - \eta  - \mu/2 ,
  \label{eq:ldifreenergy}
\end{equation}
while Eq.~\eqref{eq:density} correctly yields the density
$\rho_{\text{LD}}=1/2$. Moreover, for the HD phase, we consider
\begin{equation}
  p_{ijkl} = \delta_{i,1} \delta_{j,2} \delta_{k,1} \delta_{l,2}
\end{equation}
which, replaced into Eqs.~\eqref{eq:intenergy}
and~\eqref{eq:density}, yields respectively
\begin{equation}
  w_{\text{HD}} = 4\epsilon - 2\eta - \mu
  \label{eq:hdifreenergy}
\end{equation}
and $\rho_{\text{HD}}=1$. Finally, we have also to consider a gas
(G) phase, in which all but a finite number of sites are empty, so
that both density and energy are $0$ in the thermodynamic limit.
It is easy to show that the G phase is thermodynamically favored
($0<w_\text{LD}$ and $0<w_\text{HD}$) for $\mu <
\mu_{\text{G-LD}}$, where
\begin{equation}
  \mu_{\text{G-LD}} = 2\epsilon - 2\eta
  ,
  \label{eq:mu_G-LD}
\end{equation}
the LD phase is favored ($w_\text{LD}<0$ and
$w_\text{LD}<w_\text{HD}$) for $\mu_\text{G-LD} < \mu <
\mu_\text{LD-HD}$, where
\begin{equation}
  \mu_\text{LD-HD} = 6\epsilon - 2\eta
  ,
  \label{eq:mu_LD-HD}
\end{equation}
and the HD phase is favored ($w_{\text{HD}}<0$ and
$w_{\text{HD}}<w_{\text{LD}}$) for $\mu > \mu_{\text{LD-HD}}$. Let
us note that the LD phase has always a stability region, as
$\mu_{\text{G-LD}} < \mu_{\text{LD-HD}}$, because of the repulsive
term $\epsilon > 0$.

\section{Cluster-variation approximation}

We have performed the finite temperature analysis by means of a
cluster-variational approximation, previously applied by two of
the authors for investigating a similar water-like lattice
model~\cite{PrettiBuzano2004,PrettiBuzano2005}. The
cluster-variation method is an improved mean-field theory, which
in principle can take into account correlations at arbitrarily
large, though finite, distances. In Kikuchi's original
work~\cite{Kikuchi1951}, an approximate entropy expression was
obtained by heuristic arguments, while, in more recent and
rigorous formulations~\cite{An1988}, the approximation is shown to
be equivalent to a truncation of a cluster cumulant expansion of
the entropy. The approximation is expected to work, because of a
rapid decreasing of the cumulant magnitude, upon increasing the
cluster size, namely when the latter becomes larger than the
correlation length of the system~\cite{Morita1972}. A particular
approximation is defined by the largest clusters left in the
truncated expansion, usually denoted as basic clusters. One
obtains a free energy functional in the cluster probability
distributions, to be minimized, according to the variational
principle of statistical mechanics.

For our model we have chosen, as basic clusters, the irregular
tetrahedra which we have previously used to express the model
hamiltonian Eq.~\eqref{eq:ham}. Indeed, this choice turns out to
coincide with the (generalized) first-order approximation (on the
tetrahedron cluster), which is also equivalent to the exact
calculation for a Husimi lattice~\cite{Pretti2003}, whose
(tetrahedral) building blocks are just arranged as in
Fig.~\ref{fig:cactustetraedro}(b). The main advantage of this
approximation is its simplicity, as the only clusters retained in
the expansion are basic clusters and single sites (it is sometimes
denoted as cluster-site approximation~\cite{Oates1999}), together
with a substantial improvement over the ordinary mean-field
theory, which has been recognized for different models, even with
orientation dependent
interactions~\cite{BuzanoDestefanisPelizzolaPretti2004}. In
particular, let us note that the internal energy is treated
exactly, since the range of interactions is smaller than the basic
cluster size. The grand canonical free energy per site $\omega = w
- Ts$ ($T$ being the temperature, expressed in energy units, and
$s$ the entropy per site, in natural units) can be written as
\begin{equation}
  \frac{\omega}{T} =
  \sum_{i,j,k,l=0}^2 p_{ijkl}
  \left[ \frac{\ham_{ijkl}}{T} + \ln p_{ijkl}
  - \frac{3}{4} \ln \left( p^A_i p^B_j p^C_k p^D_l \right) \right] ,
  \label{eq:func}
\end{equation}
where $p^X_i$ is the probability of the $i$~configuration for a
site on the $X$~sublattice ($X=A,B,C,D$). These probabilities can
be obtained as marginals of the tetrahedron probability
distribution as
\begin{align}
  p^A_i & =  \sum_{j,k,l=0}^2 p_{ijkl},
  &
  p^B_j & = \sum_{k,l,i=0}^2 p_{ijkl},
  \nonumber \\
  p^C_k & =  \sum_{l,i,j=0}^2 p_{ijkl},
  &
  p^D_l & = \sum_{i,j,k=0}^2 p_{ijkl}.
  \label{eq:marginals}
\end{align}
Accordingly, the variational free energy in Eq.~\eqref{eq:func}
turns out to be a function of only the tetrahedron distribution
$p_{ijkl}$. Let us note that the free energy expression contains
two different logarithmic terms, respectively corresponding to the
cluster and site entropies. The former takes into account
correlations among four configuration variables on a tetrahedron,
while the latter can be viewed as a correction term ensuring that,
if the tetrahedron distribution factorizes into a product of
single site probabilities, the mean field (Bragg-Williams) entropy
approximation is recovered. By the way, such a heuristic argument
was first used by Guggenheim~\cite{Guggenheim1935} to derive an
equivalent expression for the case of a two-site cluster, also
equivalent to the well-known Bethe approximation~\cite{Bethe1935}.

Minimization of $\omega$ with respect to the set of tetrahedron
probabilities $\{p_{ijkl}\}$, with the normalization constraint
\begin{equation}
  \sum_{i,j,k,l=0}^2 p_{ijkl} = 1 ,
  \label{eq:constraint}
\end{equation}
can be performed by the Lagrange multiplier method, yielding the equation
\begin{equation}
  p_{ijkl} = \xi^{-1} e^{-\ham_{ijkl}/T}
  \left( p^A_i p^B_j p^C_k p^D_l  \right)^{3/4} ,
  \label{eq:cvmeq}
\end{equation}
where $\xi$ is related to the Lagrange multiplier, and can be
obtained imposing the constraint Eq.~\eqref{eq:constraint}
\begin{equation}
  \xi =
  \sum_{i,j,k,l=0}^2 e^{-\ham_{ijkl}/T}
  \left( p^A_i p^B_j p^C_k p^D_l \right)^{3/4}
  .
  \label{eq:costnorm}
\end{equation}

Eq.~\eqref{eq:cvmeq} is in a fixed point form, and can be solved
numerically by simple iteration (natural iteration
method~\cite{Kikuchi1974}). For the cluster-site approximation,
the numerical procedure can be proved to reduce the free energy at
each iteration~\cite{Kikuchi1974,Pretti2003}, and therefore to
converge to local minima. From the solution of
Eq.~\eqref{eq:cvmeq} one obtains a tetrahedron distribution
$p_{ijkl}$, whence one can compute the thermal average of every
observable; density by Eq.~\eqref{eq:density}, internal energy by
Eqs.~\eqref{eq:intenergy} and ~\eqref{eq:density}, and free energy
by Eq.~\eqref{eq:func}. The latter can be also related to the
normalization constant as~\cite{Pretti2003}
\begin{equation}
  \omega = - T \ln \xi
  .
  \label{eq:enlibcostnorm}
\end{equation}
Finally, assuming the volume per site equal to unity, pressure can
be determined, in energy units, as $P=-\omega$. In the presence of
multiple solutions, i.e., of competing phases, the
thermodynamically stable one is selected by the lowest free energy
(highest pressure) value. At zero temperature, we have $P=-w$, so
that we can easily determine the transition pressures as follows.
Replacing Eq.~\eqref{eq:mu_G-LD} into~\eqref{eq:ldifreenergy}, we
obtain the G-LD transition pressure
\begin{equation}
  P_\mathrm{G-LD} = 0 ,
\end{equation}
while, replacing Eq.~\eqref{eq:mu_LD-HD}
into~\eqref{eq:hdifreenergy}, we obtain the LD-HD transition
pressure
\begin{equation}
  P_\mathrm{LD-HD} = 2 \epsilon .
  \label{eq:p_LD-HD}
\end{equation}

\section{Results and discussion}

In this section we present and discuss our results, including a
detailed comparison with the Monte Carlo simulations of Girardi
and coworkers~\cite{GirardiBalladaresHenriquesBarbosa2007}. With
this purpose, we consider the same ratio of interaction parameters
as in Ref.~\onlinecite{GirardiBalladaresHenriquesBarbosa2007},
namely, $\eta/\epsilon=2$. Let us note that the ground-state
transition pressures computed above are independent of this
parameter.

\subsection{Phase diagram}

\begin{figure}
  \psfrag{pe}{$P/\epsilon$}
  \psfrag{te}{$T/\epsilon$}
  \includegraphics*[width=0.48\textwidth]{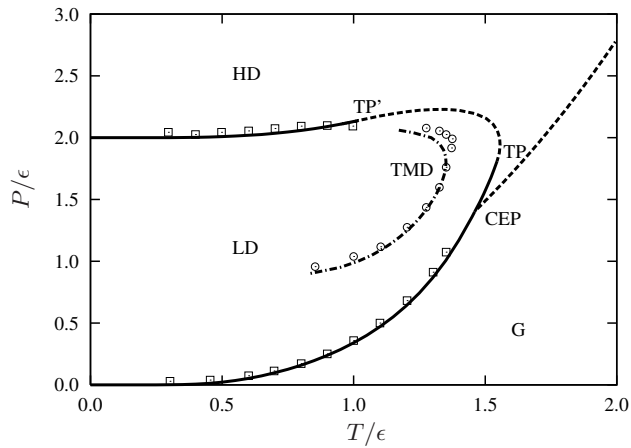}
  \caption
  {
    Pressure ($P/\epsilon$) vs temperature ($T/\epsilon$) phase diagram.
    G, LD, HD denote the corresponding phase regions (see the text);
    TP and TP' are tricritical points;
    CEP is the critical end-point.
    Solid and dashed lines denote respectively first- and second-order transitions;
    a dash-dotted line denotes the TMD locus.
    Squares and circles display Monte Carlo data from Ref.~\onlinecite{GirardiBalladaresHenriquesBarbosa2007}
    for the first-order transitions and the TMD locus, respectively.
  }
  \label{fig:pressure}
\end{figure}
\begin{figure}
  \psfrag{te}{$T/\epsilon$}
  \psfrag{rho}{$\rho$}
  \includegraphics*[width=0.48\textwidth]{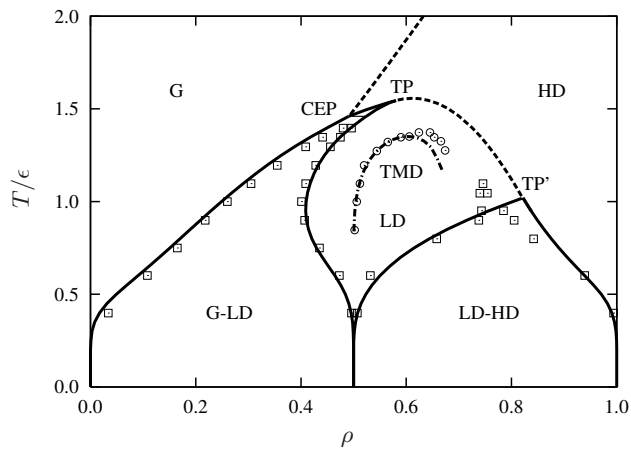}
  \caption
  {
    Temperature ($T/\epsilon$) vs density ($\rho$) phase diagram.
    Labels are defined as in Fig.~\ref{fig:pressure}
    (double labels denote coexistence regions).
    Solid lines denote boundaries of the coexistence regions,
    while dashed lines denote second-order transitions;
    a dash-dotted line denotes the TMD locus.
    Squares and circles display Monte Carlo data from Ref.~\onlinecite{GirardiBalladaresHenriquesBarbosa2007}
    for the coexistence boundaries and the TMD locus, respectively.
  }
  \label{fig:density}
\end{figure}
In Figs. \ref{fig:pressure} and~\ref{fig:density} we report two
projections of the phase diagram, respectively in the
temperature-pressure and density-temperature planes. We can
observe three different phases, which we denote as G, LD, and HD,
as they can be respectively identified as continuous evolutions of
the three ground-state phases discussed above. The G phase is
homogenous, in that the single-site probability distribution is
independent of the sublattice, whereas the LD and HD phases
exhibit a symmetry-breaking, as different sublattices behave
differently. In the homogeneous phase, molecules can locally form
H bonds, but a long-range ordered H-bond network does not appear.
Conversely, the other two phases exhibit long-range order, and one
can identify the same types of symmetry-breaking observed in the
zero-temperature LD and HD phases, respectively. More precisely,
upon decreasing temperature, the tetrahedron distributions tend to
be concentrated on the fixed configurations, corresponding
respectively to the different (LD or HD) H-bond networks described
above. For instance, in the LD phase, H bonds are preferentially
formed through $AB$ (or $BC$, $CD$, $DA$) sublattices, whereas, in
the HD phase, through $AB$ and $CD$ (or $BC$ and $DA$).

Let us have a closer look at the phase diagram topology. In the
pressure vs temperature diagram (Fig.~\ref{fig:pressure}), we
observe three different first-order transition lines. The first
one separates the G and LD phases, whereas the other two occur
between the LD and HD phases. All these lines are mapped onto
coexistence regions in the temperature vs density diagram
(Fig.~\ref{fig:density}). Both the LD-HD first-order transition
lines terminates in tricritical points (denoted as TP and TP'),
which turn out to be connected by a second-order transition line
(i.e., a line of critical points), which encloses the LD phase.
Another critical line separates the G phase from the HD phase,
which correctly prevents any continuous transformation between
these two phases, the latter having a lower symmetry. This
critical line terminates in a critical end-point (CEP). Let us
note, by the way, that, in the LD-HD coexistence region between
CEP and TP, the LD phase has in fact a higher density than HD, so
that the two phases are identified on the basis of their different
symmetries. Let us also note that the LD phase exhibits a density
anomaly, namely, a temperature of maximum density (TMD), at
constant pressure.

Figs. \ref{fig:pressure} and~\ref{fig:density} also display some
data point obtained by the Monte Carlo analysis of
Ref.~\cite{GirardiBalladaresHenriquesBarbosa2007}. We find a
remarkably good agreement both for first-order transitions
(coexistence regions in Fig.~\ref{fig:density} and transition
lines in Fig.~\ref{fig:pressure}), and for the TMD locus in the LD
phase. Deviations are slightly larger only in the vicinity of
critical lines, where the correlation length of the system
diverges, so that we expect a break-down of our approximation
performances. In spite of such an impressive agreement between
Monte Carlo results and our approximate calculation, which
suggests that the tetrahedron approximation captures the most
relevant correlations present in the system, an important
difference emerges between the two studies. Indeed, Girardi and
coworkers~\cite{GirardiBalladaresHenriquesBarbosa2007} recognize
the two long-range ordered network structures only in the
ground-state analysis. Such a long-range order seems to be lost at
finite temperature, so that they denote the two denser phases as
low density liquid (LDL) and high density liquid (HDL). In this
scenario, they only observe two first-order transitions lines
(G-LDL and LDL-HDL), terminating in two different critical points.

Although in principle such a discrepancy might be an artifact of
our approximate method, different evidences lead us to conjecture
that this is not the case, and that in
Ref.~\onlinecite{GirardiBalladaresHenriquesBarbosa2007} the
authors have simply not explored the possibility of a
symmetry-breaking. A first evidence is of course the striking
quantitative agreement shown above. Moreover, as previously
mentioned, our symmetry-broken phases at $T>0$ are continuous
evolutions of those predicted at $T=0$, whereas in
Ref.~\onlinecite{GirardiBalladaresHenriquesBarbosa2007} it is not
clear how the transitions from the zero-temperature ordered phases
to the finite temperature liquid phases occur. Looking for further
evidences, we have analyzed isotherms and isobars.

\subsection{Isotherms and isobars}

\begin{figure*}[t!]
  \psfrag{pe}{$P/\epsilon$}
  \psfrag{rho}{$\rho$}
  \psfrag{T0.8}[B,r][B,r]{$T/\epsilon = 0.8$}
  \psfrag{T1.2}[B,r][B,r]{$T/\epsilon = 1.2$}
  \psfrag{T1.6}[B,r][B,r]{$T/\epsilon = 1.6$}
  \includegraphics*[width=0.96\textwidth]{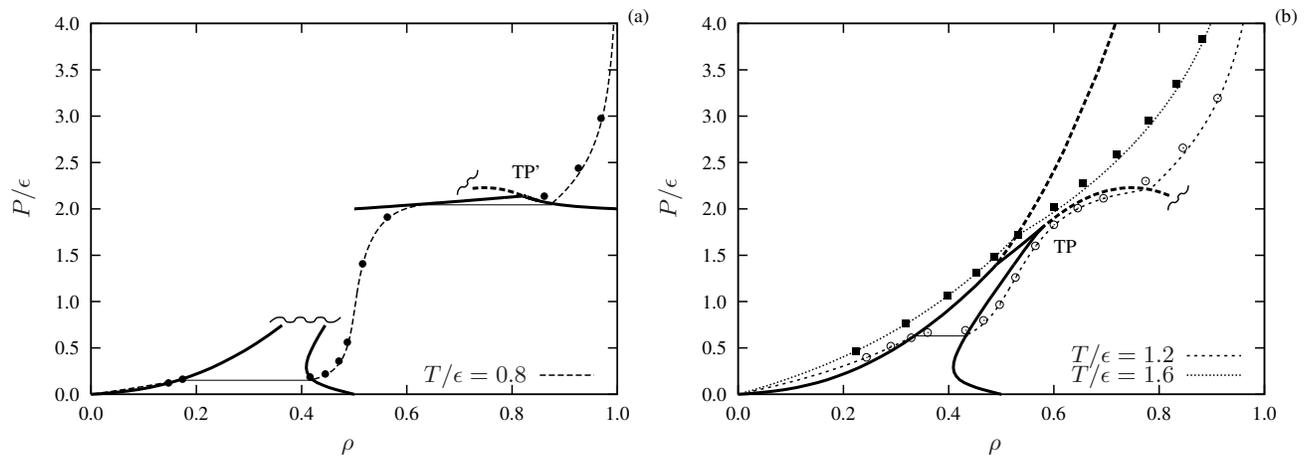}
  \caption
  {
    Isotherms in the pressure ($P/\epsilon$) vs density ($\rho$) diagram.
    Solid and dashed (thick) lines denote respectively
    boundaries of coexistence regions and critical lines
    (some lines are reported partially, to improve readability).
    Thin lines of different types represent isotherms at
    $T/\epsilon = 0.8$ (a) and $T/\epsilon = 1.2,1.6$ (b).
    Scatters denote Monte Carlo results from
    Ref.~\onlinecite{GirardiBalladaresHenriquesBarbosa2007}.
    Labels as in Fig.~\ref{fig:pressure}.
  }
  \label{fig:fix_temp}
\end{figure*}
In Fig.~\ref{fig:fix_temp} we report isotherms in a pressure vs
density diagram, together with corresponding Monte Carlo data,
available from
Ref.~\onlinecite{GirardiBalladaresHenriquesBarbosa2007}. In
particular, we consider three different temperatures, showing
qualitatively different behaviors. The first one is $T/\epsilon =
0.8$ (Fig.~\ref{fig:fix_temp}a), where the isotherm displays two
large plateaus, resulting from the two different (G-LD and LD-HD)
first-order transitions discussed above. These plateaus can be
clearly observed even from the Monte Carlo data, which are very
well fitted by our curve. The second temperature is $T/\epsilon =
1.2$ (Fig.~\ref{fig:fix_temp}b), lying between the two tricritical
points TP and TP' (see Figs. \ref{fig:pressure}
and~\ref{fig:density}). In this case, upon increasing pressure,
the system first undergoes a G-LD first-order transition, and then
a LD-HD second-order transition. The former results in a plateau
as well, whereas the latter gives rise to a simple ``kink'' in the
isotherm. The Monte Carlo data exhibit a similar plateau and,
noticeably, they also seem to be compatible with the possibility
of a critical-line crossing, i.e., with the existence of a kink in
the thermodynamic limit. Finally, we consider $T/\epsilon = 1.6$
(Fig.~\ref{fig:fix_temp}b), lying above both tricritical points.
Here the system directly undergoes a (second-order) G-HD
transition, so that the isotherm displays just a slight kink. Even
in this case, the slope change observed in the Monte Carlo
isotherm seems to be well explained by the critical-line crossing.

Let us note that, looking at Figs. \ref{fig:pressure}
and~\ref{fig:density}, we indeed expect two more qualitatively
different regimes, respectively corresponding to the narrow
temperature intervals between CEP and TP and between TP and the
maximum temperature reached along the TP-TP' critical line. In the
former case, upon increasing pressure, we find a G-HD second-order
transition, followed by a HD-LD first-order transition, and
finally by a LD-HD second-order transition. In the latter case, we
find the same sequence of transitions, all second-order.
Corresponding isotherms are not displayed in
Fig.~\ref{fig:fix_temp}. Indeed, we suppose that a comparison with
Monte Carlo simulations may be very difficult in these temperature
region, where maximum quantitative discrepancies are observed in
the phase diagram (see Fig.~\ref{fig:density}). To be more
precise, it is likely that similar behaviors might also appear
from Monte Carlo data, but at a slightly lower temperature than
observed from the approximate calculation.

\begin{figure*}
  \psfrag{rho}{$\rho$}
  \psfrag{P2.0}[B,r][B,r]{$P/\epsilon = 2.0$}
  \psfrag{P1.8}[B,r][B,r]{$P/\epsilon = 1.8$}
  \psfrag{P1.6}[B,r][B,r]{$P/\epsilon = 1.6$}
  \psfrag{P1.4}[B,r][B,r]{$P/\epsilon = 1.4$}
  \psfrag{P1.2}[B,r][B,r]{$P/\epsilon = 1.2$}
  \psfrag{P1.0}[B,r][B,r]{$P/\epsilon = 1.0$}
  \psfrag{P2.07}[B,r][B,r]{$P/\epsilon = 2.07$}
  \psfrag{P1.83}[B,r][B,r]{$P/\epsilon = 1.83$}
  \psfrag{P1.35}[B,r][B,r]{$P/\epsilon = 1.35$}
  \psfrag{P0.87}[B,r][B,r]{$P/\epsilon = 0.87$}
  \psfrag{tempe}{$T/\epsilon$}
  \includegraphics*[width=0.96\textwidth]{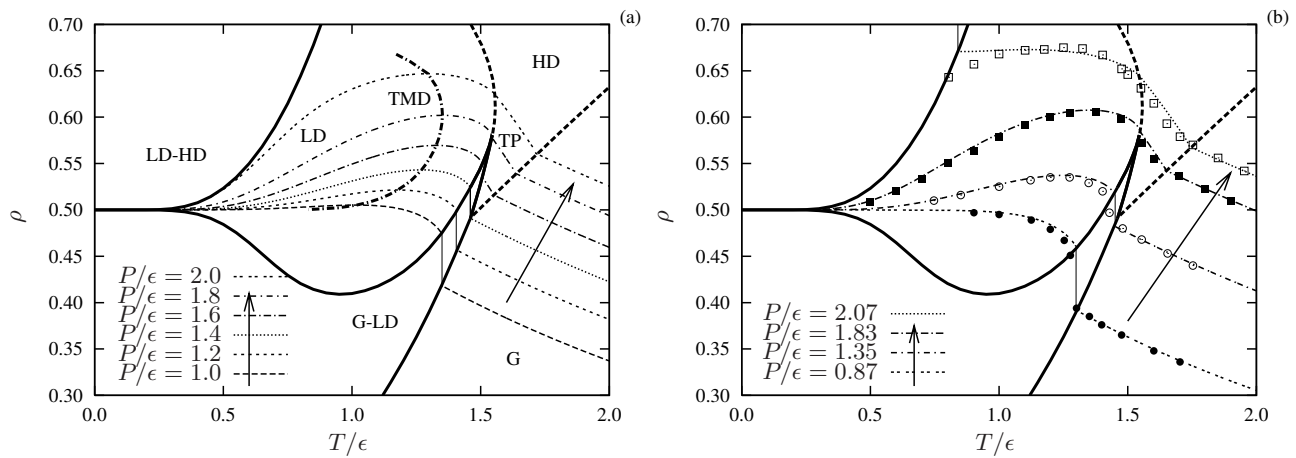}
  \caption
  {
    Isobars in the density ($\rho$) vs temperature ($T/\epsilon$) diagram.
    Solid and dashed (thick) lines denote respectively
    boundaries of coexistence regions and critical lines.
    Thin lines of different types represent isobars at various pressures.
    In (a), a thick dash-dotted line denotes the TMD locus.
    In (b), scatters denote Monte Carlo results from
    Ref.~\onlinecite{GirardiBalladaresHenriquesBarbosa2007}.
    Labels as in Fig.~\ref{fig:density}.
  }
  \label{fig:fix_press}
\end{figure*}
In Fig.~\ref{fig:fix_press} we report isobars in a density vs
temperature diagram (more precisely, Fig.~\ref{fig:fix_press}a
displays only results from our calculation, whereas in
Fig.~\ref{fig:fix_press}b some Monte Carlo results are
superimposed). We have considered different pressure values, below
the TP' pressure, pointing out four different regimes. The lowest
pressure isobars lie below the CEP pressure
$P_\mathrm{CEP}/\epsilon \approx 1.4$, so that the system only
undergoes a first-order G-LD transition, upon decreasing
temperature (see Fig.~\ref{fig:pressure}). At the lowest
pressures, the density maximum turns out to be extremely shallow,
but it becomes more pronounced upon increasing pressure (the TMD
locus correctly connects all density maxima). At pressure values
lying between $P_\mathrm{CEP}$ and the TP pressure
$P_\mathrm{TP}/\epsilon \approx 1.8$, the system phase behavior
becomes more complicated. Upon decreasing temperature, we first
observe a second-order G-HD transition followed by a weak
first-order HD-LD transition. The HD-LD transition also becomes
continuous above $P_\mathrm{TP}$. Finally at $P/\epsilon>2$ [i.e.,
above the ground-state LD-HD transition pressure, see
Eq.~\eqref{eq:p_LD-HD}], a further LD-HD first-order transition
appears at low temperature. A comparison of these results with
Monte Carlo isobars drawn from
Ref.~\onlinecite{GirardiBalladaresHenriquesBarbosa2007} generally
confirms the good agreement obtained for phase diagrams and
isotherms. In particular, let us note that the G-HD critical line
provides a clear interpretation of the slope changes observed in
the two highest pressure isobars in Fig.~\ref{fig:fix_press}b.

Looking at Fig.~\ref{fig:pressure}, we expect two more regimes,
for pressure values above TP'. More precisely, between TP' and the
maximum pressure value reached along the TP-TP' critical line, we
observe a sequence of three continuous transitions, namely, G-HD,
HD-LD, and LD-HD, upon decreasing temperature. For higher
pressure, we finally observe a direct G-HD continuous transition.
Even in these cases, corresponding isobars have not been reported
in Fig.~\ref{fig:fix_press}.

\section{A Monte Carlo test}

As pointed out in the previous section, different evidences
support the existence of long-range ordered symmetry-broken
phases, and related critical transitions. Nevertheless, due to the
approximate nature of our results, we have found it necessary to
perform some specific checks by Monte Carlo simulations. We have
only investigated certain regions around which the transitions are
expected to appear.

In order to verify the G-HD transition, we consider a constant
temperature $T/\epsilon = 1.7$ (see Figs. \ref{fig:density}
and~\ref{fig:pressure}), and work out a grand-canonical simulation
for varying chemical potential. The lattice is made up of $L
\times L \times L$ bcc conventional cells (see
Fig.~\ref{fig:reticolo}), with $L$ ranging from 10 up to 30, and
periodic boundary conditions. The simulation is of the Metropolis
type, starting from a random lattice configuration. We randomly
select a lattice site, and flip its configuration according to the
Metropolis rule. We define the Monte Carlo step as $2L^3$
iterations of this process. Measurements are performed every
30-100 Monte Carlo steps, depending on the phase (the Monte Carlo
dynamics is faster in the disordered G phase than in the LD and HD
phases). We collect up to $10^5$ measurements, waiting a
thermalization time of order $10^3$ measurements.

\begin{figure}
  \psfrag{cumulant}{$V_w$}
  \psfrag{L10}[B,r][B,r]{$L = 10$}
  \psfrag{L14}[B,r][B,r]{$L = 14$}
  \psfrag{L20}[B,r][B,r]{$L = 20$}
  \psfrag{L30}[B,r][B,r]{$L = 30$}
  \psfrag{mue}{$\mu/\epsilon$}
 \includegraphics*[width=0.48\textwidth]{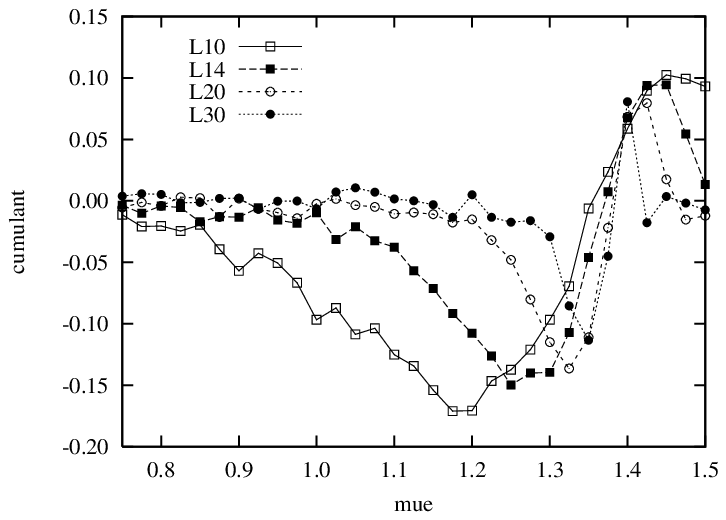}
 \caption
 {
   Fourth-order cumulant of the grand canonical energy ($V_w$)
   vs chemical potential ($\mu/\epsilon$)
   at constant temperature $T/\epsilon = 1.7$,
   for different values of the lattice size $L$.
   Error bars, not reported for a better readability,
   are of order $10^{-2}$ in the worst case.
 }
 \label{fig:critical}
\end{figure}
In Fig.~\ref{fig:critical} we report the fourth-order 
cumulant~\cite{Binder1981} of
the grand-canonical energy $V_w$ as a function of the chemical
potential, for different values of the lattice size $L$. We can
observe that $V_w$ fluctuates around zero, except in a narrow
interval (which shrinks upon increasing~$L$) where $V_w$ exhibits
a minimum followed by a maximum. This is a typical signature of a
second-order transition~\cite{TsaiSalinas1998}, which in our case
we expect to be the G-HD transition. Actually, in both phases far
from the transition, the energy distribution can be described by a
single peak, which can be approximated by a gaussian, tending to a
Dirac delta in the $L \rightarrow \infty$ limit. In this case
$V_w$ should vanish independently of~$L$. Upon approaching the
critical line, the symmetric gaussian peak description is no
longer valid, so that $V_w \neq 0$~\cite{TsaiSalinas1998}. Though
we do not report the details here, we have worked out an
equivalent analysis for the LD-HD critical line, obtaining similar
results.

The results displayed in Fig.~\ref{fig:critical} seem to indicate
a transition around $1.3 \lesssim \mu/\epsilon \lesssim 1.4$,
where the simulation predicts a density $0.59 \lesssim \rho
\lesssim 0.61$. The latter values turn out to be slightly higher
than the transition density predicted by the cluster-variation
approximation $\rho \approx 0.56$, at the corresponding transition
chemical potential $\mu/\epsilon \approx 1.00$. We then expect
that the actual critical line is shifted towards higher density
(i.e., higher pressure) with respect to the approximate line shown
in Figs. \ref{fig:pressure} and~\ref{fig:density}. This result is
in agreement with the known behavior of cluster-variational
approximations, which, due to their mean-field nature, usually
overestimate the ordered phase region.

In order to better characterize the transition, we define the
following order parameter
\begin{equation}
 \phi_1 = p_1^{AC} - p_1^{BD}
 ,
 \label{eq:symm}
\end{equation}
where $p_1^{XY}$ is the probability of the $i=1$ configuration
(see Fig.~\ref{fig:molecole}) on the $X$ and $Y$ sublattices. A
totally equivalent order parameter $\phi_2$ could be obtained
considering the $i=2$ configuration. Let us note that, as
previously mentioned, the HD phase is two-fold degenerate.
Therefore, we expect that our simulated system can go into one of
the two possible states, depending on initial conditions. In one
case, preferential H bonding occurs either through $AB$ and $CD$
sublattices, with preferred $i=1$ configurations on $A$ and $C$.
In the other case, preferential bonding occurs through $BC$ and
$DA$, with preferred $i=1$ configurations on $B$ and $D$. In the
two cases we respectively obtain positive and negative values of
the order parameter.

\begin{figure}
  \psfrag{phiL}{$|\phi_1|$}
  \psfrag{L10}[B,r][B,r]{$L = 10$}
  \psfrag{L14}[B,r][B,r]{$L = 14$}
  \psfrag{L20}[B,r][B,r]{$L = 20$}
  \psfrag{L30}[B,r][B,r]{$L = 30$}
  \psfrag{mue}{$\mu/\epsilon$}
 \includegraphics*[width=0.48\textwidth]{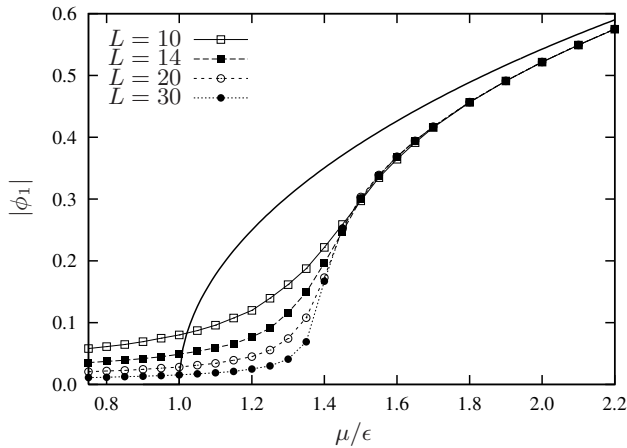}
 \caption
 {
   G-HD order parameter ($|\phi_1|$) vs chemical potential ($\mu/\epsilon$)
   at constant temperature $T/\epsilon = 1.7$.
   The solid line denotes the cluster-variation result,
   while scatters denote Monte Carlo results
   for different values of the lattice size $L$
   (thin lines are an eyeguide).
   Error bars are of order $10^{-4}$.
 }
 \label{fig:symm}
\end{figure}
In Fig.~\ref{fig:symm} we report the absolute value of $\phi_1$,
computed both by simulations (for different lattice sizes~$L$) and
by the cluster-variation approximation (in the thermodynamic
limit), as a function of the chemical potential. In the latter
case, we compute the two-sublattice probabilities as
\begin{equation}
  p_i^{XY} = \frac{p_i^X+p_i^Y}{2} ,
\end{equation}
where the one-sublattice probabilities are derived from
Eqs.~\eqref{eq:marginals}. It is easy to see that $\phi_1$ behaves
like a good order parameter for the G-HD transition. Indeed, for
low $\mu$ values, where we expect to observe the disordered G
phase, $\phi_1$ tends to zero upon increasing~$L$. Conversely, for
higher $\mu$ values, where we expect to observe the ordered HD
phase, $\phi_1$ remains nonzero, with a smaller effect of~$L$. In
the intermediate region, the slope of the curves increases upon
increasing $L$, revealing the phase transition. Let us note, by
the way, that for $L=10$ (the size used in
Ref.~\onlinecite{GirardiBalladaresHenriquesBarbosa2007}) finite
size effects are definitely not negligible in the vicinity of the
transition, where the correlation length of the system is expected
to diverge. Accordingly, also the cluster-variational result fails
in this region, although it progressively improves upon moving
away from the transition.

\section{Conclusions}

In this paper, we have considered a lattice model with
orientation-dependent interactions, meant to describe a so-called
network-forming fluid. As mentioned in the Introduction, this
model is an instance of quite a large class of similar models
(defined on the body-centered cubic lattice, with tetrahedral
model molecules), originally conceived to describe water and
investigate its various anomalies. Several studies have been
carried out using a variety of approximate techniques, whose
reliability has not always been checked. Motivated by a recent
work by Girardi and
coworkers~\cite{GirardiBalladaresHenriquesBarbosa2007}, performing
extensive Monte Carlo simulations for one of such models, we have
worked out a corresponding analysis by means of a
cluster-variational technique, with the aim of comparing the
results and evaluating the reliability of the approximate method.
The latter, based on a tetrahedral cluster, had been already
employed by two authors of the present paper for investigating a
different water-like lattice model~\cite{PrettiBuzano2004}, and
actually turns out to be a generalization of the original
first-order approximation worked out by Bell~\cite{Bell1972}. The
generalization consists in taking into account the possibility of
a symmetry-breaking at the level of four face-centered cubic
sublattices of the original body-centered cubic lattice. Such a
symmetry breaking is indeed known to occur in the model
ground-state.

As already pointed out in the text, we obtain a remarkably good
agreement with Monte Carlo results for most thermodynamic
quantities. Nevertheless, our results clearly suggest that the the
two denser phases, identified as liquid in
Ref.~\onlinecite{GirardiBalladaresHenriquesBarbosa2007}, are in
fact endowed with a long range ordered structure, closely
resembling the hydrogen bond networks observed at zero
temperature. Accordingly, the emerging picture of the phase
diagram turns out to be much more complex and richer. In fact, due
to symmetry-breaking, the two critical points observed in
Ref.~\onlinecite{GirardiBalladaresHenriquesBarbosa2007} turns out
to be tricritical, and two new critical lines emerge, separating
the two symmetry-broken phases from each other and from the
disordered phase. Such critical lines are hardly detectable by
observing macroscopically averaged quantities (such as the
density), since they manifest themselves only as kinks (not
plateaus) in isotherms and isobars, which are further smoothed out
by finite size effects. We have indeed performed a Monte Carlo
simulation, in order to check the qualitative correctness of our
results, around the critical lines determined by the approximate
method. The density exhibits in fact a smooth behavior, but the
analysis of a Binder cumulant~\cite{TsaiSalinas1998}, and of a
suitable order parameter, shows clear evidence of an
order-disorder critical transition.

As previously mentioned, these results suggest that the
tetrahedral cluster approximation is able to capture the most
relevant correlations present in the system. On the other hand,
unfortunately, the observed phase diagram seems to make the model
less relevant for investigating water. Concerning this issue, the
main problem is of course that the two ordered phases, which
indeed may well represent two different ice
forms~\cite{BellSalt1976}, undergo second-order transitions, which
have never been observed experimentally. It is likely that further
variations of the model might improve the similarity of its phase
diagram to that of real water, but a discussion of this point is
beyond the scope of the present paper and is left for future work.


\end{document}